\documentclass[10pt,letterpaper,twocolumn]{article} 

\usepackage{ol2}
\usepackage[draft]{hyperref}
\usepackage{amsmath}

\begin{document}

\twocolumn[ 

\title{The Maryland model in optical waveguide lattices}


\author{Stefano Longhi}
\address{Dipartimento di Fisica, Politecnico di Milano and Istituto di Fotonica e Nanotecnologie del Consiglio Nazionale delle Ricerche, Piazza L. da Vinci 32, I-20133 Milano, Italy (stefano.longhi@polimi.it)}
\address{IFISC (UIB-CSIC), Instituto de Fisica Interdisciplinar y Sistemas Complejos, E-07122 Palma de Mallorca, Spain}

\begin{abstract}
The Maryland model was introduced more than 30 years ago  as an integrable model of localization by aperiodic order.
Even though quite popular and rich of fascinating mathematical properties,
this model has so far remained quite artificial, as compared to other models displaying dynamical localization like the periodically-kicked quantum rotator or the Aubry-Andre$^{\prime}$ model.
 Here we suggest that light propagation in a polygonal optical waveguide lattice provides a photonic realization of the Maryland model and enables to observe a main prediction of this model, namely fragility of wave localization in the commensurate potential limit.
\end{abstract}

 ] 


{\it Introduction.} Optical waveguide lattices with disorder or aperiodic order have provided a fantastic platform to observe with integrated photonics groundbreaking physical phenomena like Anderson localization, metal-insulator phase transitions and Anderson topological phases \cite{r1,r2,r3,r4,r5,r6,r7,r8,r9,r10,r11}, with potential applications to the design of photonic and plasmonic devices \cite{r12,r13,r14,r15}. 
Notably, arrays of evanescently-coupled waveguides have been used to emulate a popular model of aperiodic order in integer quantum Hall systems, the Aubry-Andre$^{\prime}$-Harper model \cite{r2,r3}. Another famous model of Anderson-like localization in one-dimensional incommensurate potentials was introduced by Grempel, Fishman, and Prange \cite{r16} in connection with the problem of quantum chaos and  dynamical localization in periodically-kicked quantum systems \cite{r17,r18,r19}. This exactly-solvable model was dubbed the Maryland model by Barry Simon \cite{r20,r21}, who studied in great details its mathematical properties revealing fascinating and unusual features deeply rooted in number theory. In Ref.\cite{r22}, it was shown that the Maryland model represents a topological quantum phase transition point
in a class of corresponding two-dimensional lattice models with integer quantum Hall topology, thus connecting the Maryland model to the broad area of topological phases of matter.
 While the periodically-kicked quantum rotator model has been implemented in different classical and quantum systems \cite{r23,r24,r25,r26,r27,r28,r29}, the Maryland model has remained so far a rather artificial model. The reason thereof is that, while in the quantum rotator model the kinetic energy operator is represented by the physical quadratic term of particle momentum, in the Maryland model the kinetic energy term should depend linearly on particle momentum, which has seemed to be an artificial mathematical assumption to make the model integrable.\par
In this Letter we suggest an optical system, based on light propagation in a polygonal array of evanescently-coupled optical waveguides, which realizes the Maryland model and that should be of easy experimental implementation with current integrated photonic technologies.\\
\\
{\it Maryland model.}  
Following the original paper by Grempel, Fishman, and Prange \cite{r16}, let us consider the dynamics of a periodically-kicked quantum particle by a spatially periodic potential, with a kinetic energy operator which is a linear function of particle momentum. The dynamics is described by the Schr\"odinger equation in dimensionless form
\begin{equation}
i \frac{\partial \Psi}{\partial t}=K (\hat{p}_x) \Psi + V(x) \sum_n \delta(t-n) \Psi
\end{equation}
\begin{figure}[htb]
\centerline{\includegraphics[width=8.4cm]{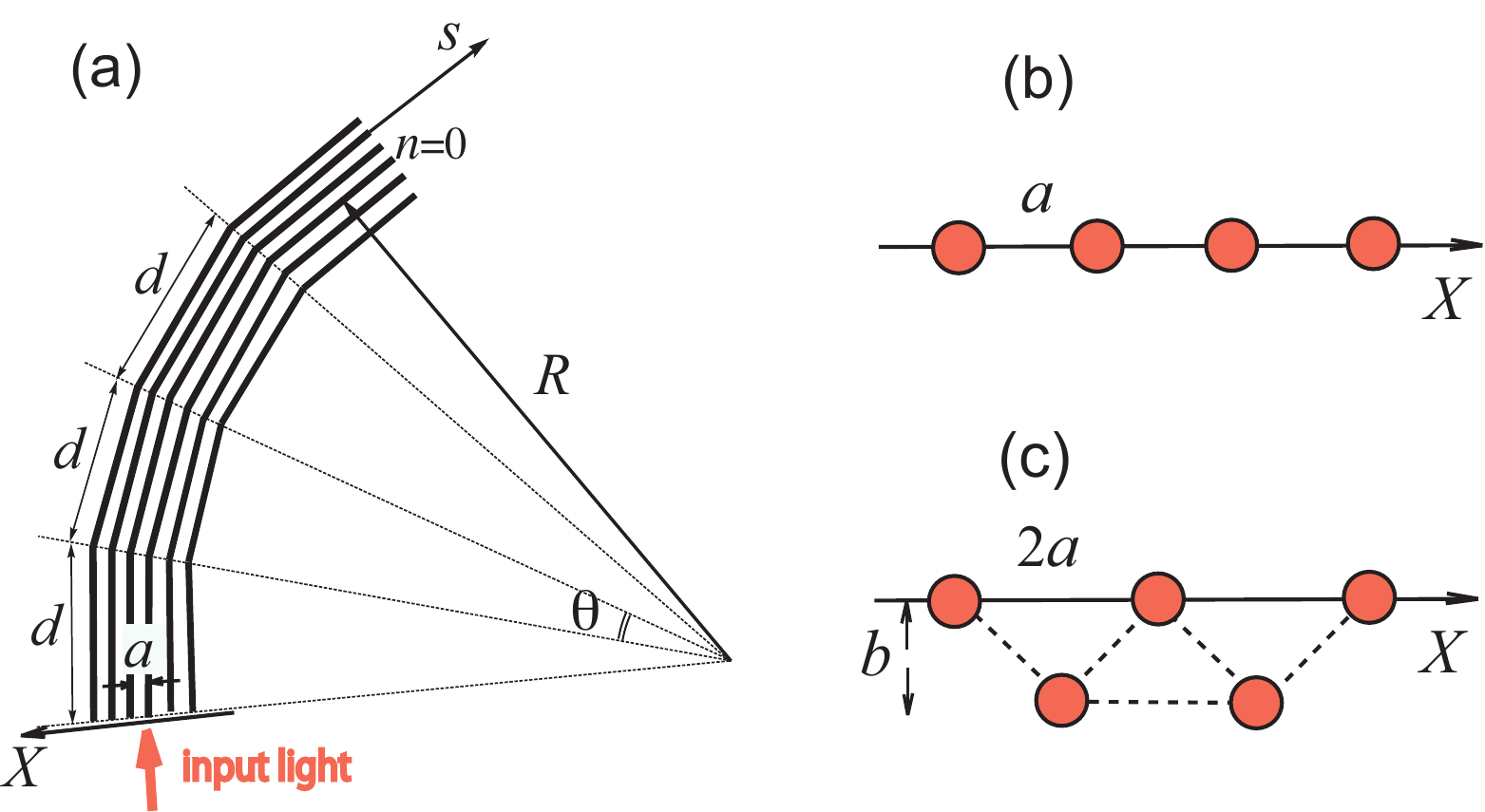}} \caption{ 
(Color online) (a) Schematic of a polygonal waveguide array. A lattice of optical guides with period $a$ is bent along a polygonal line $s$ with a small tilt angle $\theta$ every distance $d=R \theta$, with $R \gg a$. The radius $R$ is taken at the $n=0$ reference waveguide in the lattice. 
(b,c) Coupling constants between waveguides in the lattice can be controlled by transverse geometric setting. In the linear geometry of (b) the dominant term is the nearest-neighbor hopping term $\Delta_1$, while in the zig-zag geometry of (c) the 
nearest-neighbor hopping term $\Delta_1$ and next-to-the nearest-neighbor hopping term $\Delta_2$ are comparable.}
\end{figure} 
for the wave function $\Psi=\Psi(x,t)$, where $\hat{p}_x= -i \partial_x$ is the particle momentum operator, $K(p_x)= 2 \pi \alpha p_x$ is the linear dispersion relation of the kinetic energy, and $V(x)=V(x+2\pi)$ is the external period potential with $ 2 \pi$ spatial period. 
Applying standard Floquet theory, after setting $\Psi(x,t)=u(x,t) \exp(-i \mu t)$, where $-\pi \leq \mu < \pi$ is the quasi-energy and $u(x,t+1)=u(x,t)$ is the periodic part of the wave function, the following equation is readily found
\begin{equation}
\exp(-i \mu) u(x)=\exp [-i V(x)] \exp [-i K(\hat{p}_x)] u(x)
\end{equation}
for the function $u(x) \equiv u(x,0^+)$. After setting  $\psi(x)=u(x)/[1-i W(x)]$, with $W(x)={ \rm tan} [V(x)/2]$, from Eq.(2) it readily follows that the Fourier coefficients $ \psi_n$ of $\psi(x)$ satisfy the spectral problem \cite{r16,r17,r18}
\begin{equation}
\sum_{l \neq n }W_{n-l} \psi_l+ { \rm tan} \left( \pi \alpha n- \frac{\mu}{2} \right) \psi_n=E \psi_n
\end{equation}
where $W_n$ are the Fourier coefficients of $W(x)$, i.e. $W(x)=\sum_n W_n \exp(in x)$, and $E=-W_0$. The spectral problem defined by Eq.(3), dubbed the Maryland model by B. Simon \cite{r20}, is integrable \cite{r16,r20}. The general result is that, for almost every irrational $\alpha$, the energy spectrum is pure point and the eigenfunctions are exponentially localized. This means that any initially-localized excitation does not spread in momentum (Fourier) space and remains localized, an effect dubbed dynamical localization. On the other hand, for rational $\alpha$ the spectrum is absolutely continuous, formed by many closely-spaced energy bands, the wave functions are extended (Bloch-type), and dynamical localization in momentum space is rather generally prevented. Interestingly, in Ref.\cite{r16} it was predicted that, if $V(x)$ is of sinusoidal shape, i.e. its Fourier spectrum is formed solely by the three harmonics $V_{0, \pm 1}$, then even for $ \alpha$ rational one observes localization owing to band flattening. However, such a localization effect for $\alpha$ rational is fragile, since even small higher-order Fourier terms of the potential $V(x)$ break exact band flattening and enable wave diffusion in momentum space.\\
\\
{\it  Polygonal waveguide lattice.}  The key observation to implement the Maryland model in waveguide lattices is to note that the Floquet spectral problem defined by Eq.(2) can be likewise obtained from the modified Schr\"odinger equation 
\begin{equation}
i \frac{\partial \Psi}{\partial t}= K(\hat{p}_x) \sum_n \delta(t-n) \Psi + V(x) \Psi 
\end{equation} 
which differs from Eq.(1) because the kicks occur now  on the kinetic energy operator  $K(\hat{p}_x)$, rather than on the potential $V(x)$. In fact,  applying standard Floquet theory we set $\Psi(x,t)=u(x,t) \exp(-i \mu t)$ in Eq.(4), were $-\pi \leq \mu < \pi$ is the quasi-energy and $u(x,t+1)=u(x,t)$ is the periodic part of the wave function. Then it can be readily shown that $u(x) \equiv u(x,1^-)$ satisfies again the spectral equation (2), and thus the dynamics described by Eq.(4) displays in momentum space the same localization/delocalization features than the Maryland model. To emulate the periodically-kicked system described by Eq.(4), let us consider propagation of monochromatic light waves in a waveguide lattice with lattice period $a$ and with an optical axis $s$ periodically bent by a small angle $\theta$ at successive intervals, spaced by $d=R \theta$, to form an open polygonal line of radius $R \gg a$, as schematically shown in Fig.1(a). Light propagation in the waveguide lattice,  along the polygonal abscissa  $s$, is governed by the following coupled-mode equations for the modal field amplitudes $c_n(s)$ in the various guides \cite{r5,r30,r31}
\begin{equation}
i \frac{d c_n}{ds}= \sum_{l \neq n} \Delta_{n-l} c_l+ 2 \pi \alpha n c_n \sum_{l} \delta (s- l d)
\end{equation}
 where $\Delta_{l}=\Delta_{-l}$ is the coupling constant between waveguides, distant by $l$ lattice sites, and 
 \begin{equation}
 \alpha= \frac{n_s  \theta a }{\lambda}
 \end{equation}
is the phase gradient introduced by the axis bent \cite{r31}. In Eq.(6), $ \lambda$ is the wavelength (in vacuum) of the probing light and $n_s$ the effective waveguide mode index.
The coupling constants $\Delta_l$ largely depend on the geometrical setting of waveguides in the transverse plane \cite{Dreisow}, with almost nearest-neighbor couplings in the straight geometry of Fig.1(b) 
($\Delta_l \neq 0$ only for $l= \pm 1$) and controllable next-to-nearest neighbor couplings in the zig-zag geometry of Fig.1(c) ($\Delta_l=0$ for $l \neq \pm 1, \pm2$).
 We note that, in the continuous limit
$ \theta, d \rightarrow 0$ with $R=d / \theta$ finite, i.e. when the polygonal line becomes a circumference of radius $R \gg a$ and the discrete nature of periodic phase kicks is smeared out, Eq.(5) describes Bloch oscillations in circularly curved waveguide lattices, demonstrated in previous works \cite{r32b,r32,r33}. In this continuous limit localization is always ensured by the formation of a Wannier-Stark ladder energy spectrum, regardless the parameter $\alpha$ is a rational or irrational number and for arbitrary long-range hopping. Conversely, in the polygonal array setup of Fig.1(a) with periodic phase gradient kicks, it can be readily shown that light propagation reproduces the Maryland model, and thus the rational/irrational value of $\alpha$ makes the difference. In fact, after setting $t=s/d$ and $\Psi(x,t)= \sum_n c_n(t) \exp(i x n)$, from Eq.(5) it follows that $\Psi(x,t)$ satisfies Eq.(4) with 
\begin{equation}
V(x)= d  \sum_l \Delta_l \exp(i l x)
\end{equation}
 and $K(p_x)=2 \pi \alpha p_x$. The absence of dynamical localization for rational $\alpha$ can be physically understood by observing that the effect of each kick on a Bloch wave, $c_n \sim \exp(iqn)$, is to shift the Bloch wave number from $q$ to $q^{\prime}=q-2 \pi \alpha$. For $\alpha$ rational,  there is a finite set of Bloch wave numbers that are invariant under the change $q \rightarrow q+2 \pi \alpha$, so that the eigenfunctions of the periodic system can be formed by a suitable superposition of such Bloch waves, resulting in extended wave functions and thus delocalization.\\
  \\
 \begin{figure}[htb]
\centerline{\includegraphics[width=8.7cm]{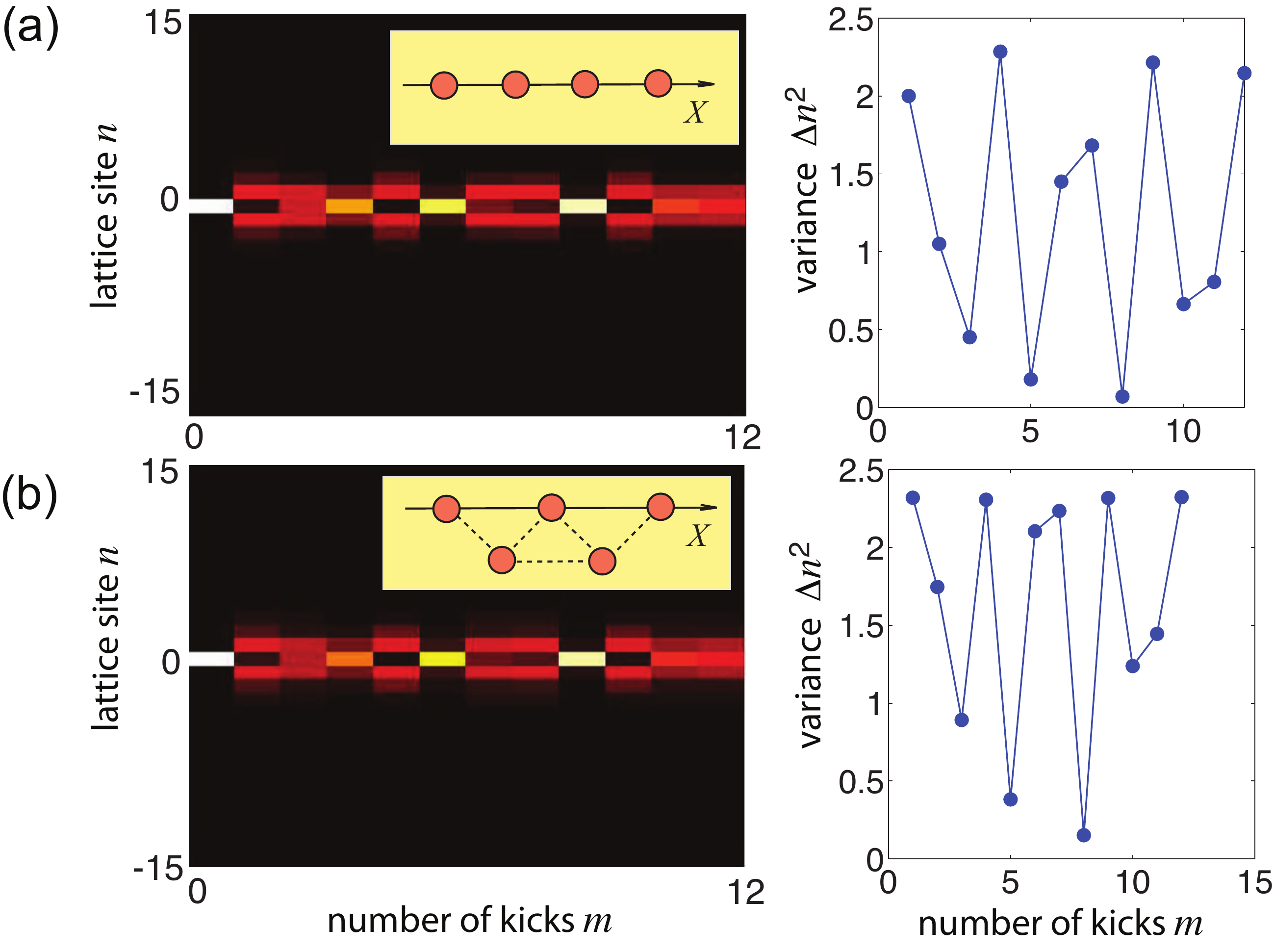}} \caption{ 
(Color online) Dynamical localization in the polygonal waveguide lattice in the incommensurate case $\alpha=(\sqrt{5}-1)/2$. The figure depicts the numerically-computed  light intensity distribution on a pseudo color map in the waveguide lattice at subsequent kicks $m=s/d$ (left panels) and corresponding variance $\Delta n^2=\sum_n n^2 |c_n|^2$ of the distribution (right panels).  Light is initially injected into waveguide $n=0$. In (a) we consider a linear geometry with only nearest neighbor coupling $\Delta_1$, while in (b) we consider a zig-zag geometry with next-to-the-nearest neighbor coupling $\Delta_2= 0.2 \Delta_1$. The spatial distance $d$ between two consecutive kicks is set equal to $d=1 / \Delta_1$.}
\end{figure}
 {\it Dynamical localization.} According to the spectral properties of the Maryland model \cite{r16,r20,r21}, light  spreading in the waveguide lattice of Fig.1(a) is suppressed for almost every irrational $\alpha$ (dynamical localization), while rather generally wave spreading can be observed for rational values of $\alpha$. For an array with nearest-neighbor couplings solely, dynamical localization also arises due to band flattening \cite{r16} and the formation of a Wannier-Stark ladder similar to what happens in the Bloch oscillation problem with circularly-curved waveguides. In fact, for $\Delta_l=0$ when $l \neq \pm 1$, Eq.(5) can be exactly solved using, for example, the method described in \cite{r31}. In particular, let us assume that at the input plane $s=0$ the waveguide $n=0$ is initially excited, which corresponds to the simplest experimental condition of array excitation. From the solution to Eq.(5) with the initial condition $c_{n}(0)=\delta_{n,0}$, one obtains the following expression for the light intensity $|c_n (s=md)|^2$ at the $n$-th waveguide in the lattice and at the $m$-th kick  
  \begin{equation}
 |c_n(s=md)|^2= J_n^2  \left( 2 \Delta_1 d \left|  \sum_{l=0}^{m-1} \exp(2 \pi i \alpha l) \right|  \right)
 \end{equation}
 \begin{figure}[htb]
\centerline{\includegraphics[width=8.7cm]{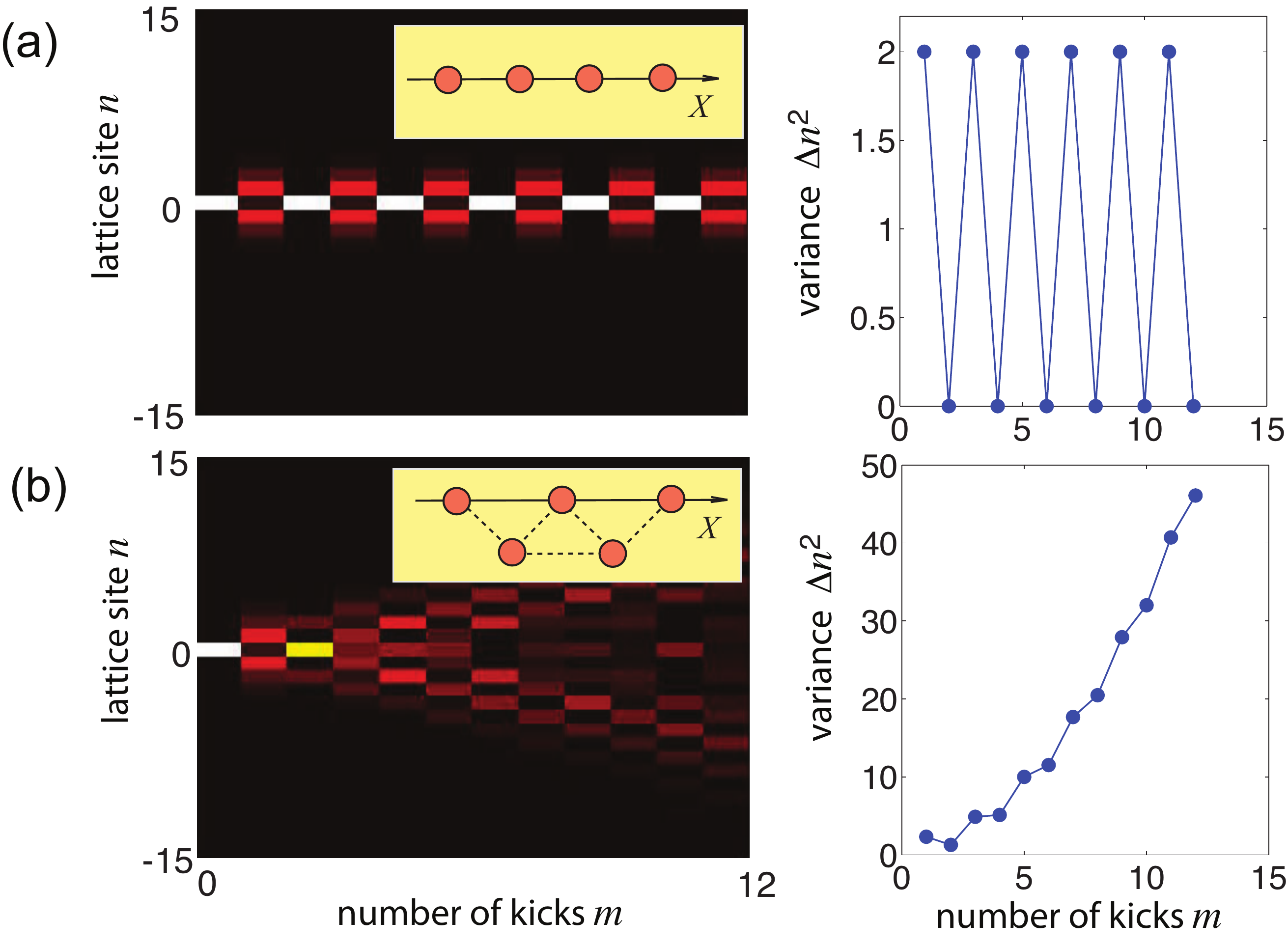}} \caption{ 
(Color online) Same as Fig.2 but for the commensurate case $\alpha=1/2$. Note that in this case delocalization is observed in the zig-zag geometry [panel (b)].}
\end{figure} 
where $J_n$ are the Bessel functions of first kind.  Note that, since the geometric progression $\sum_{l=0}^{m-1} \exp(2 \pi i \alpha l)$ in the argument function on the right hand side of Eq.(8) remains bounded as the number of kicks $m$ increases, light diffusion in the lattice is prevented for any (either rational or irrational) value of $\alpha$. However, while for almost every irrational $\alpha$ the dynamical localization is a robust effect, it becomes fragile for $\alpha$ rational \cite{r16}.\\ 
 We have checked the predictions of the theoretical analysis by direct numerical simulations of coupled-mode equations (5), considering two representative geometrical settings: a linear array [Figs.1(b)], where the coupling is almost limited to nearest-neighbor guides, and a zig-zag array [Fig.1(c)], where second-order coupling is non-negligible \cite{Dreisow}.
 \begin{figure}[htb]
\centerline{\includegraphics[width=8.7cm]{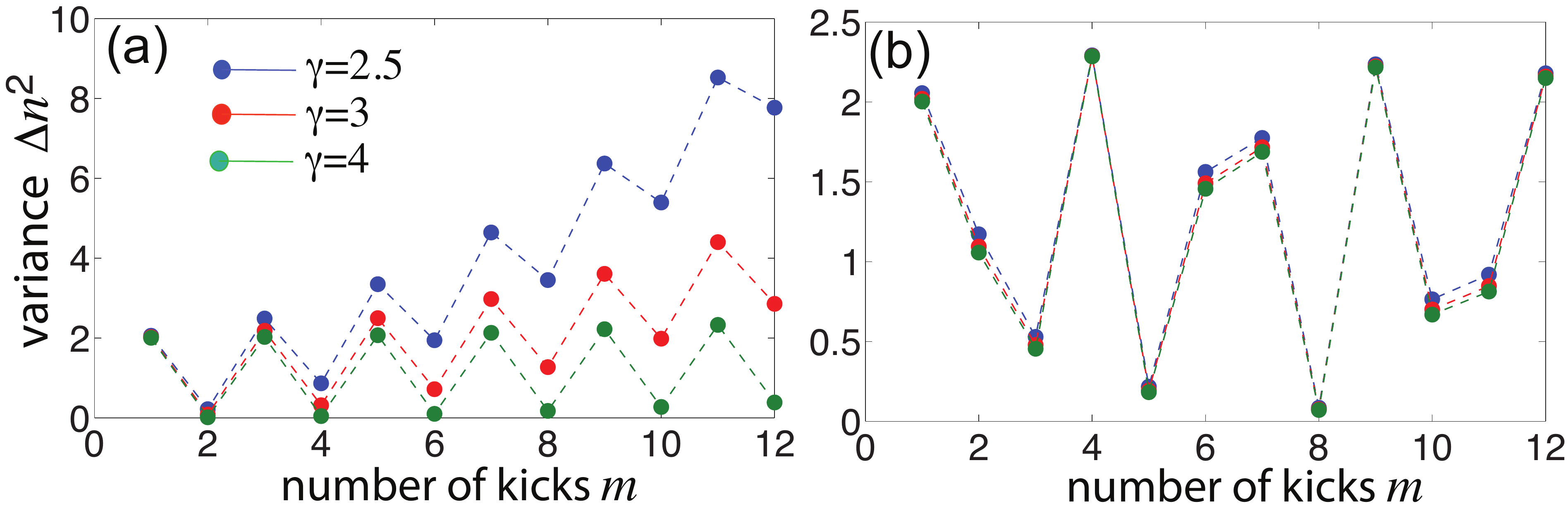}} \caption{ 
(Color online) Behavior of light beam variance $\Delta n^2$ versus kick number $m$ is a linear array with long-range couplings $\Delta_l= \Delta_1 \exp(-\gamma |l|+\gamma)$ for a few increasing values of $\gamma$. In (a) $\alpha=1/2$, in (b) $\alpha= (\sqrt{5}-1)/2$.}
\end{figure}
 \begin{figure}[htb]
\centerline{\includegraphics[width=8.7cm]{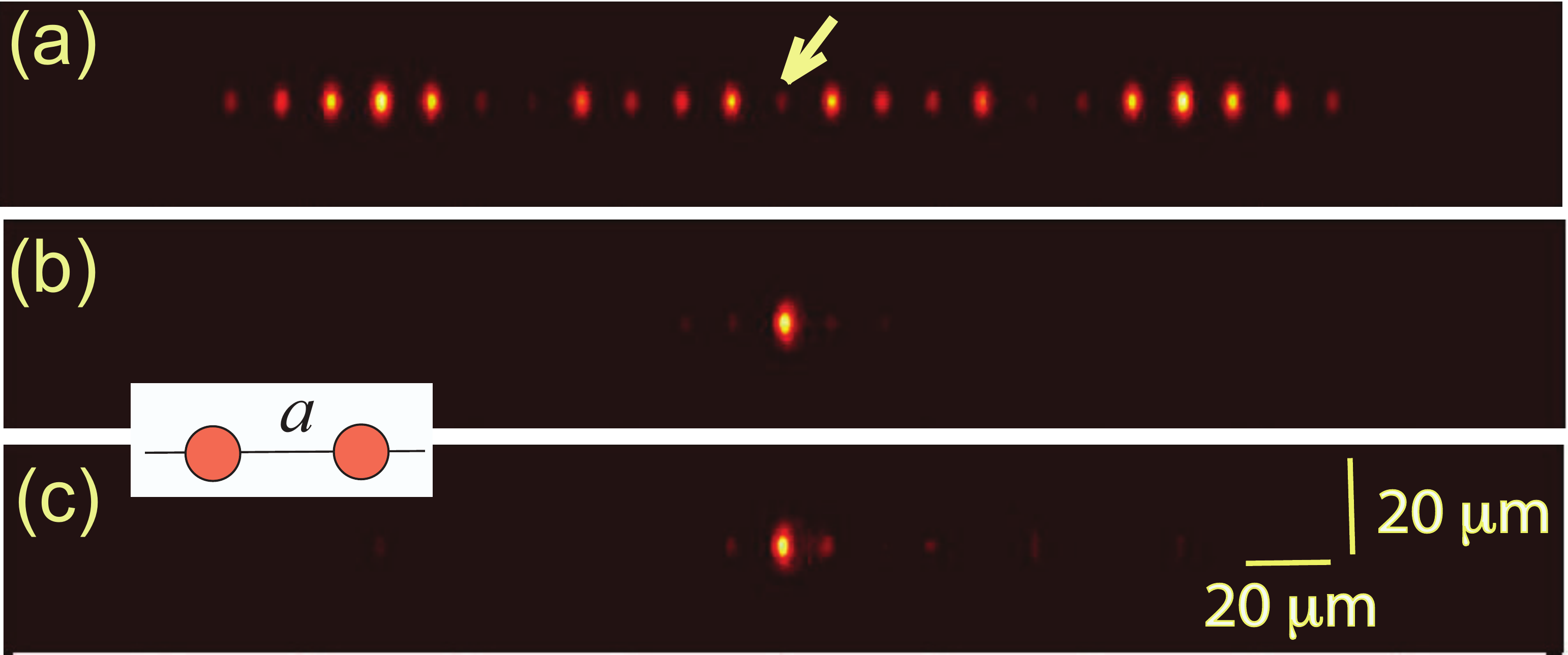}} \caption{ 
(Color online) Light intensity distribution at the output plane of a $L=6$-cm-long linear waveguide array (waveguide spacing $a=16 \; \mu$m), probed with red light ($\lambda=633$ nm) for 
(a) $\alpha=0$ (no phase gradients), (b) $\alpha=0.5$, and (c) $\alpha=(\sqrt{5}-1)/2$. The distance between two consecutive kicks is $d=7$ mm. The arrow shows the waveguide excited at the input plane.}
\end{figure}
 \begin{figure}[htb]
\centerline{\includegraphics[width=8.7cm]{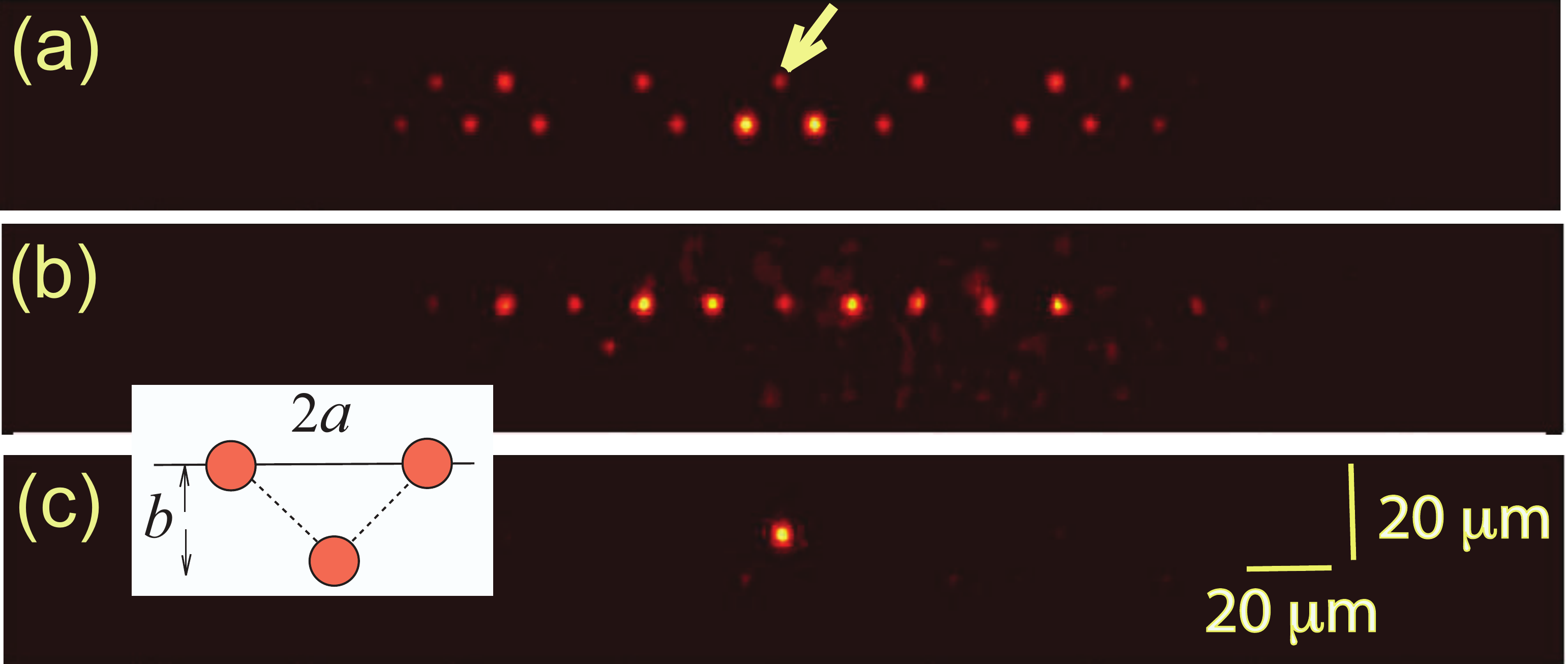}} \caption{ 
(Color online) Same as Fig.5 but for a zig-zag waveguide array. Waveguide spacings are $a=11 \; \mu$m and $b=12 \; \mu$m.}
\end{figure}
 An example of dynamical localization for incommensurate $\alpha=(\sqrt{5}-1)/2 \simeq 0.618$ (the inverse of golden ratio) is shown in Fig.2. The waveguide $n=0$ in the array is excited at the input plane $s=0$, and the evolution of the discretized light intensity is displayed at successive kicks, i.e. at $s=0,d,2d,3d,...$. Note that in both geometrical settings wave spreading in the lattice is suppressed. Figure 3 shows the same behavior for a rational $\alpha=1/2$. Note that in this case dynamical localization is not observed anymore in the zig-zag geometrical setting, involving second-order couplings [Fig.3(b)]. It should be noted that in real waveguide arrays the coupling constant between two waveguides is a nearly exponential decaying function of waveguide spacing \cite{r30,Dreisow}, so that strictly speaking the residual long-range couplings 
 yield delocalization for $\alpha$ rational also in the linear geometry of Fig.1(b), However, since the decay of $\Delta_{l}$ with $|l|$ is typically fast, delocalization is not visible for short propagation distances (kick numbers) accessible in a typical experiment. This is illustrated in Fig.4, which shows the behavior of the variance $\Delta n^2$ versus kick  number $m$ (up to 12 kicks) in a linear array with long-range hopping $\Delta_{l}=\Delta_1 \exp(-\gamma |l|+\gamma)$ for a few decreasing values of $\gamma$.\\ 
To check the validity of coupled-mode theory, we numerically simulated light propagation in the waveguide lattice by solving the three-dimensional optical Schr\"odinger equation \cite{r5} using a standard pseudo spectral split-step method. In the simulations, we assumed a circular profile of the guide core with a super-Gaussian profile of radius $2 \; \mu m$, a peak refractive index change $\Delta n=0.0015$, and a substrate refractive index $n_s=1.5$ at the probing wavelength $\lambda=633$ nm (red light). Such parameter values are typical of waveguide arrays manufactures by the femtosecond (fs) laser writing technique in fused silica\cite{r10,r30,Dreisow,r32,r33}. A sample length $L=6$ cm is assumed, with a sequence of waveguide axis tilt (periodic phase kicks) spaced by $d=7$ mm, corresponding to a total number of 8 kicks. Figures 5 and 6 show the light intensity distribution at the output plane of the array, i.e. after a propagation distance $s=L=6$ cm, for the two geometrical settings of linear array (Fig.5) and zig-zag array (Fig.6). The numerical results clearly show that, while in the linear array light remains trapped  for both $\alpha$ rational and irrational, in the zig-zag array dynamical localization is fragile for $\alpha$ rational, resulting in a consistent beam broadening [Fig.6(b)].\\
\\		 
{\it Conclusion.} We suggested an integrated photonic system, based on light propagation in a polygonal waveguide lattice, which realizes a famous integrable model of localization, the Maryland model. The setup can reveal the fragility of dynamical localization in the commensurate potential limit.
 Our results should be feasible for an experimental observation with current integrated-optic technology, and are expected to be of relevance in different areas of physics beyond photonics.\\

\end{document}